\documentclass[twocolumn,showkeys,superscriptaddress,preprintnumbers,amssymb,floatfix,nofootinbib]{revtex4}
\usepackage{epsfig,graphics}
\usepackage{graphicx}
\usepackage{dcolumn}
\usepackage{bm}
\usepackage{amsmath}
\usepackage[usenames]{color}
\usepackage{ulem} 
\usepackage{booktabs}
\usepackage{array}  
\usepackage{CJK}
\usepackage[colorlinks,linkcolor=blue,urlcolor=blue,citecolor=blue]{hyperref}
\usepackage{tikz}
\usepackage{csquotes}
\usepackage{hyperref}
\usetikzlibrary {shapes.geometric}
\usetikzlibrary{shapes,arrows}
\begin{document}
\preprint{RIKEN-iTHEMS-Report-25}
\begin{CJK*}{UTF8}{gbsn}

\title{Neural Unfolding of the Chiral Magnetic Effect in Heavy-Ion Collisions}
\author{Shuang Guo (郭爽)}
\affiliation{Key Laboratory of Nuclear Physics and Ion-beam Application (MOE), Institute of Modern Physics, Fudan University, Shanghai 200433, China}
\affiliation{Shanghai Research Center for Theoretical Nuclear Physics, NSFC and Fudan University, Shanghai $200438$, China}
\author{Lingxiao Wang (王凌霄)}
\email[]{lingxiao.wang@riken.jp}
\affiliation{RIKEN Interdisciplinary Theoretical and Mathematical Sciences (iTHEMS), Wako, Saitama 351-0198, Japan}
\author{Kai Zhou (周凯)}
\email[]{zhoukai@cuhk.edu.cn}
\affiliation{School of Science and Engineering, The Chinese University of Hong Kong, Shenzhen (CUHK-Shenzhen), Guangdong, $518172$, China}
\author{Guo-Liang Ma (马国亮)}
\email[]{glma@fudan.edu.cn}
\affiliation{Key Laboratory of Nuclear Physics and Ion-beam Application (MOE), Institute of Modern Physics, Fudan University, Shanghai 200433, China}
\affiliation{Shanghai Research Center for Theoretical Nuclear Physics, NSFC and Fudan University, Shanghai $200438$, China}

\begin{abstract}

The search for the chiral magnetic effect (CME) in relativistic heavy-ion collisions (HICs) is challenged by significant background contamination. We present a novel deep learning approach based on a U-Net architecture to time-reversely unfold the dynamics of CME-related charge separation, enabling the reconstruction of the physics signal across the entire evolution of HICs. Trained on the events simulated by a multi-phase transport model with different cases of CME settings, our model learns to recover the charge separation based on final-state transverse momentum distributions at either the quark-gloun plasma freeze-out or hadronic freeze-out. This devises a methodological tool for the study of CME and underscores the promise of deep learning approaches in retrieving physics signals in HICs.

\end{abstract}



\maketitle

\section{INTRODUCTION}
\label{introduction}

Topological fluctuations of gluon fields in the quantum-chromodynamic (QCD) vacuum can induce a chirality imbalance, which implies local parity violation in strong interactions~\cite{QCD1,QCD2,QCD3}. This imbalance, quantified by a chiral chemical potential ($\mu_5$), generates an electric current $\mathbf{J}$ parallel to an external magnetic field $\mathbf{B}$, resulting in a dipole-like charge separation (CS) along $\mathbf{B}$---a phenomenon known as the chiral magnetic effect (CME)~\cite{CME1,CME2,CME3,CME4,CME5}. Relativistic heavy-ion collisions provide a unique experimental platform for studying the CME, as they simultaneously create the high-temperature QCD environment required for chirality imbalance and generate extremely strong magnetic fields in non-central collisions~\cite{magentic,Bzdak:2011yy,early_stage,Chen:2021nxs,Xu:2014tda,Tang:2011xq,Zhao:2017rpf,Zhao:2022dac,Gao:2020vbh,Wu:2021xgu,Shou:2024uga,Shen:2025unr}. Consequently, extensive experimental searches for CME-induced charge separation have been carried out at both the Relativistic Heavy Ion Collider (RHIC) at Brookhaven National Laboratory (BNL)~\cite{RHIC1,RHIC2,RHIC3,Chen:2024aom} and the Large Hadron Collider (LHC) at CERN~\cite{LHC1,LHC2}.

However, since the magnetic field exists only transiently during the early stages of the collision~\cite{magentic, Bzdak:2011yy, early_stage, Chen:2021nxs} and the late-stage evolution of heavy-ion collisions (HICs) generates significant background sources (e.g., elliptic flow), the detection of the CME faces substantial challenges~\cite{background1,background2,background3, background4}. To extract potential CME signals from final-state hadrons, various analysis techniques and observables have been developed~\cite{CME_analysis1,CME_analysis2,CME_analysis3}. A notable example is the CME-motivated isobar collision program conducted at RHIC in 2018~\cite{isobar1,isobar2}. However, another major obstacle arises from substantial background contributions, particularly due to the difference in nuclear structure between two isobaric nuclei ~\cite{background_nuclear_structure1,background_nuclear_structure2, background_nuclear_structure3}. As a result, none of the current methods can unambiguously identify the CME or its induced charge separation (CS) in the quark-gluon plasma (QGP) along the direction of the magnetic field in isobar collisions. Given that event statistics in Au+Au collisions at 200 GeV will increase by a factor of 10 or more by the end of 2025, more advanced analysis methods are required to search for the potential CME signal. Promising techniques include event shape engineering~\cite{event_shape_engneering}, event shape selection~\cite{event_shape_selection1,event_shape_selection2} and spectator/participant plane methods~\cite{spectator_participant_plane1,spectator_participant_plane2,Chen:2023jhx,Chen:2025jmb}.

At the same time, the rapid advancement of computational hardware and machine learning (ML) algorithms has revolutionized the way to extract information from large-scale datasets. Deep learning (DL), as a specialized ML paradigm, demonstrates exceptional capability in identifying nonlinear patterns with complex correlations~\cite{DL1,DL2}. Its hierarchical neural architectures and automated feature extraction mechanisms are particularly effective in modeling complex physical systems, finding widespread applications in nuclear physics~\cite{DL_nuclear1,DL_nuclear2,DL_nuclear3,DL_nuclear4,DL_nuclear5,pointnet2,DL_nuclear7,DL_nuclear8,DL_nuclear9,DL_nuclear10,Zhou:2023pti}, particle physics~\cite{DL_particle1,DL_particle2,DL_particle3,DL_particle4,DL_particle5}, and condensed matter physics~\cite{DL_condensed1,DL_condensed2,DL_condensed3,DL_condensed4}.

In the DL study of the CME, supervised learning approach with convolutional neural networks has demonstrated success in identifying charge separation signals within final-state pion spectra~\cite{CME_meter}. However, this approach faced two significant challenges: first, the CME signal undergoes substantial changes during the dynamical evolution of HICs~\cite{AMPT_CS,AVFD_1,CKT_Sun_1}; second, flow-related backgrounds progressively dominate as the heavy-ion collision system evolves~\cite{background2}. Developing DL techniques to reconstruct the evolution of CME signals in the QGP based on the measured final-state hadrons could provide a powerful solution to address these challenges, potentially overcoming current limitations in CME detection. In this Letter, we propose a novel deep learning framework based on a multi-phase transport (AMPT) model to unfold the final-state CME signal and reconstruct its complete evolution throughout the entire HICs via time reversal.

This paper is organized as follows: Section~\ref{AMPT} introduces the AMPT model which provides the training and test datasets, while discussing its key model parameters, initial conditions, and their relevance to the CME. Section~\ref{U-net} presents the time-embedded U-Net architecture, detailing its encoder-decoder structure with skip connections that enables accurate reconstruction of CME signal evolutions. In Sec.~\ref{results}, we present three cases with distinct inputs, demonstrating the predicted event-average charge separation (CS) along the reverse time evolution and the CS distribution at each time step comparing with the ground truth values.  Finally, we summarize and discuss the implications of our results in Sec.~\ref{summary}.

\section{MODEL AND METHOD}
\label{model}
 \subsection{A multi-phase transport model}
\label{AMPT}

The AMPT model used in this work incorporates the string melting mechanism~\cite{AMPT}, simulating the complete evolution of high-energy heavy-ion collisions through four important stages. The initial conditions are provided by the HIJING (Heavy Ion Jet INteraction Generator) model~\cite{HIJING1,HIJING2}, which generates the spatial and momentum distributions of minijet partons from hard processes and soft string excitations from soft processes. In the string melting implementation, all excited strings of hadrons are converted to their constituent partons while preserving the valence quarks' flavor and spin structures~\cite{HIJING3}. The produced partons undergo scattering only after a string melting formation time $t_f$,
\begin{equation}\label{t_f}
    t_f =\frac{E_H}{m_{T,H}^2}
\end{equation} 
where $E_H$ and $m_{T,H}$ represent the energy and transverse mass of the parent hadrons, respectively.
The initial positions of partons originating from melted strings are determined by tracing their parent hadrons along straight-line trajectories. 
The interactions among partons inside the formed QGP are described by Zhang's parton cascade (ZPC) model \cite{ZPC}, which includes only two-body elastic scatterings with a g+g $\rightarrow $ g+g cross section, i.e.,
\begin{equation}\label{sigma}
\frac{d\sigma }{d\hat{t}}=\frac{9\pi \alpha ^{2}_{s}}{2}(1+\frac{\mu^2 }{\hat{s}})\frac{1}{(\hat{t}-\mu^2)^2}
\end{equation}
where $\alpha_s$ is the strong coupling constant (taken as 0.33), while $\hat{s}$ and $\hat{t}$ are the usual Mandelstam variables. The effective screening mass $\mu$ is taken as a parameter in ZPC to adjust the parton scattering cross section. In our study, we adopt a parton scattering cross section of $\sigma=3$ mb, a value well established for describing experimental observables at both RHIC and LHC energies~\cite{3mb0,3mb1,3mb2,3mb3,3mb4,3mb5}. To systematically track the evolution of the QGP, we record partonic data at intervals $\delta t = 0.2\,\mathrm{fm}/c$, starting from the initial state until parton freeze-out. At the QGP freeze-out, the AMPT model undergoes the transition from the partonic phase to the hadronic phase by recombining partons into hadrons using a naive nearest quark coalescence mechanism~\cite{AMPT}. Finally, the hadronic scatterings in the hadronic phase are simulated by a relativistic transport (ART) model \cite{ART}.

The original AMPT model lacks an intrinsic generation mechanism of chiral magnetic effect (CME). We implemented a global charge separation (CS) scheme following Ref.~\cite{AMPT_CS}, where we interchange a fraction of the $p_y$ values between downward-moving $u$ quarks and upward-moving $\bar{u}$ quarks (and similarly for $\bar{d}/d$ and $\bar{s}/s$ quark pairs) to simulate CME-like initial CS conditions. Here, \enquote{upward} and \enquote{downward} are defined relative to the $y$-axis, which is perpendicular to the reaction plane. The CS fraction $f$ is defined as~\footnote{The CS fraction definition can be directly extended to characterize hadronic matter, when N is the number of charged hadrons.},
\begin{equation}
    f=\frac{N^{\pm}_{\uparrow  (\downarrow)}-N^{\pm}_{\downarrow  (\uparrow)}}{N^{\pm}_{\uparrow  (\downarrow)}+N^{\pm}_{\downarrow  (\uparrow)}}
\label{charge_separation}
\end{equation}
where $N$ is the number of quarks of a given species, $+$ and $-$ denote positive and negative charges, respectively, and $\uparrow$ while $\downarrow$ represent the moving directions. Here we use a right-handed coordinate system with the x-axis along the impact parameter $b$, the z-axis along the beam direction, and the y-axis along the magnetic field direction completing the orthogonal set.

\begin{figure*}[hbtp]
	\includegraphics[scale=0.30]{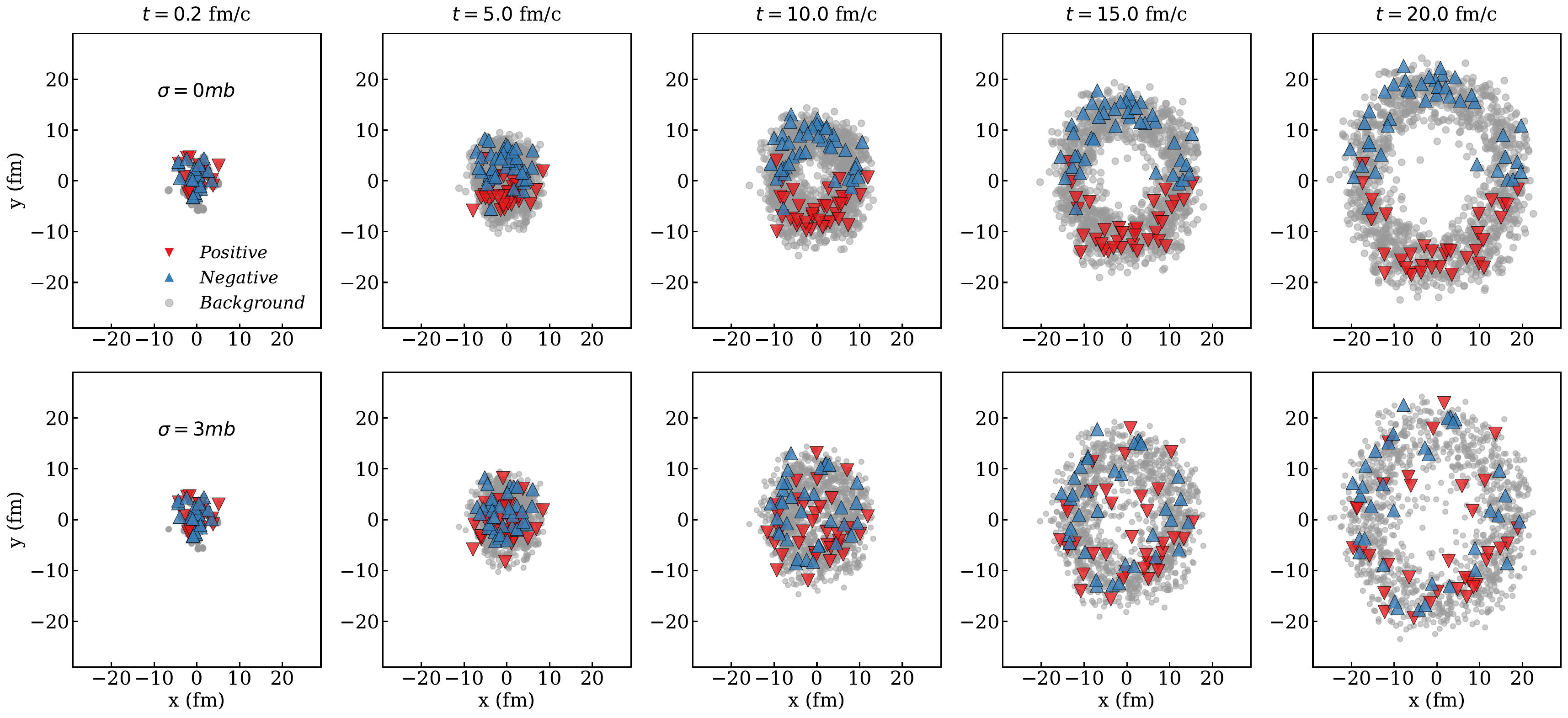}
	\caption{The evolution of the CME signal in the QGP medium, displayed in the transverse $X$-$Y$ plane at different times: $t = 0.2, 5.0, 10.0, 15.0$, and $20.0 \,\mathrm{fm}/c$. The upper panels show the non-interacting scenario ($\sigma = 0$ mb), while the lower panels present the case with partonic interactions ($\sigma = 3$ mb). The CME signal partons are indicated by triangular markers (two colors denotes positive and negative charges), with the background QGP partons represented by gray circles.}
	\label{xyevolution}
\end{figure*}

Figure~\ref{xyevolution} demonstrates the evolution of charge separation (CS) between positively and negatively charged signal partons in the QGP, highlighting the crucial role of partonic interactions. The CME-signal partons are represented by red (positive) and blue (negative) triangles, indicating quarks with initially swapped $p_y$ momenta in the modified AMPT framework, while gray circles depict the background partons. The upper panel presents the non-interacting scenario ($\sigma = 0$\,mb), where positive and negative signal partons maintain clearly separated spatial distributions with the evolution. In striking contrast, the lower panel for finite partonic cross-section ($\sigma = 3$\,mb) shows progressive degradation of the initial CS configuration, manifested through increasing spatial overlap between oppositely charged signal partons. This comparison provides direct evidence that partonic interactions strongly suppress the CS development, leading to significant reduction of the CME signal. Quantitative analysis reveals that final-state interactions suppress an initial CS fraction of 10\% down to 1--2\% in Au+Au collisions at 200 GeV~\cite{AMPT_CS,Zhao:2022grq}.

In this study, we simulate Au+Au collisions at $\sqrt{s_{NN}}$ = 200 GeV with a fixed impact parameter of $b$ = 8 fm, representative of the 30-40\% centrality bin, the main focus of the CME search. To investigate the dynamics of CME-induced charge separation (CS), we first establish a baseline with an initial CS fraction of $f = 10\%$ and further vary $f$ across a broad range from $1\%$ to $10\%$ in $1\%$ increments to simulate the real fluctuation effect to enhance our approach's robustness. The QGP evolution is simulated using our modified AMPT model incorporating the CME effect, with a fine temporal resolution of $0.2\,\mathrm{fm}/c$ to accurately track the dynamical evolution. The simulation includes both the hadronization process and subsequent hadronic freeze-out stage. Our central objective is to time-reversely recover CME dynamics through deep learning (DL) analysis of late-stage information, including from the QGP freeze-out and final-state hadrons. The reconstruction faces two major challenges: (1) the degradation of initial CS due to partonic interactions in the QGP phase, and (2) further signal dilution during hadronization and hadronic rescatterings. To address these challenges, we will employ deep learning techniques to effectively extract and reconstruct the CME signal from the complex final-state data.

\subsection{The deep learning architecture U-Net}
\label{U-net}
U-Net, a widely used convolutional neural network model, is particularly effective due to its symmetric encoder-decoder structure and skip connections~\cite{Unet_arti}.

\begin{figure*}[hbtp]
	\includegraphics[trim=100 90 30 50, clip,scale=0.5]{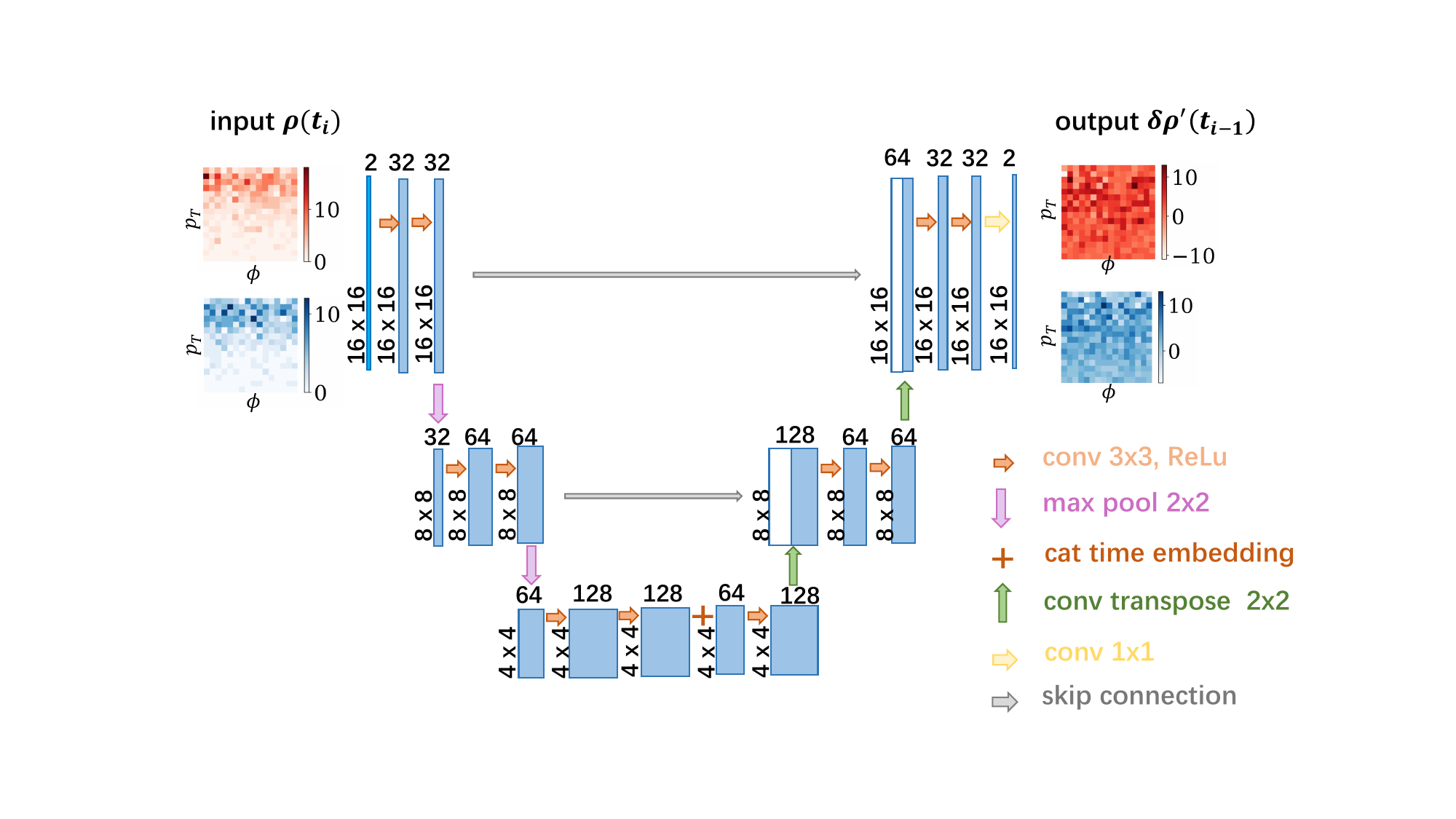}
	\caption{Schematic illustration of the U-Net architecture used in our analysis. Each blue box corresponds to a multi-channel feature map, annotated with channel numbers (top) and spatial dimensions (bottom left). The white boxes represent copied feature maps. The arrows denote the different operations. The network processes two-channel input representing the transverse momentum distribution $(p_T, \phi)$ of final-state particles, with positive and negative charges encoded as separate channels (red and blue colormaps, respectively). The output predicts the difference between distributions at adjacent time steps.}
	\label{unet}
\end{figure*}

The architecture of the neural network we used is shown in Figure~\ref{unet}, which is a U-Net-based architecture designed for pattern retrieval tasks, integrating the time embedding technique to enhance temporal awareness. It consists of an encoder-decoder structure with skip connections, where the encoder comprises multiple convolutional layers with $3\times 3$ kernels and ReLU activation, followed by $2\times2$ max-pooling layers for downsampling. The decoder mirrors the encoder with $2\times2$ transposed convolutions for upsampling, effectively restoring spatial resolution while utilizing skip connections to preserve spatial details. A key feature of our model is the incorporation of time embeddings, which are concatenated at the bottleneck stage to provide temporal context. Due to the special architecture, the U-Net is particularly suitable for capturing time-dependent phase space evolution.

The input to the U-Net model is the transverse momentum $(p_T, \phi)$-distribution of final-state partons or hadrons, represented as a two-dimensional image $\rho({t_i})$ at a time step $i$ ($i=n,n-1,...,3,2,1$). The supervised learning target at each step $i$ is the change in the distribution between two adjacent time steps or stages, defined as, 
\begin{equation}\label{output} \delta\rho(t_{i-1}) = \rho(t_{i-1}) - \rho(t_i),
\end{equation} while the model is trained to predict this difference as $\delta\rho'(t_{i-1})$.
Here, $\rho({t_{i}}) $ denotes the true distribution at the current time step, and $\rho({t_{i-1}})$ corresponds to the distribution at the previous time step or stage. By using the difference between two adjacent time steps as the learning target, the model is trained to capture the temporal evolution of the distribution. The predicted difference $\delta \rho'({t_{i-1}})$ is then added to the input $\rho(t_i)$ to obtain the reconstructed distribution at the previous step,
\begin{equation}\label{input}
 \rho'(t_{i-1}) =\delta\rho'({t_{i-1}}) + \rho(t_i).
\end{equation}
 This reconstructed distribution $\rho'(t_{i-1}) $ is used as the input for the next iteration. Training the model to predict these differences helps it effectively capture the reverse evolution of the system at each pixel. This iterative framework enables progressive reconstruction of the charge separation (CS), allowing the trained model to infer earlier distributions from any given time step. To ensure accurate reconstruction, the loss function consists of the Mean Squared Error (MSE) between the target and predicted distributions, supplemented by the MSE of their CS to preserve the dynamics of CME.

For data preparation, we follow the STAR method as described in Ref.~\cite{pt_cut}, implementing kinematic cuts with pseudorapidity $|\eta| <$ 1.0 and transverse momentum 0.15 $< p_{\rm T} <$ 2 GeV/\textit{c} for charged hadrons, 0.075 $< p_{\rm T} <$ 1 GeV/\textit{c} for partons. We utilize the transverse momentum-space particle number density distribution, represented as a 16 × 16 pixel image, for either partonic or hadronic final states as input to the U-Net model. Particles are categorized into positive and negative charges, which are treated as two separate input channels. Each channel contains a $16 \times 16$ image representing the transverse momentum distribution of particles with a given charge. We generate a dataset consisting of 5,000 events for training , which is confirmed to achieve stable convergence and consistent
performance, and 10,000 events for testing to ensure a more robust evaluation.

The encoder captures hierarchical spatial features of the input distribution, while the time embeddings provide essential temporal context, enabling the model to differentiate between various evolution stages of the system. The decoder, equipped with upsampling layers and skip connections, preserves fine-grained spatial details throughout the reconstruction process. This structured architecture allows the model to effectively learn the dynamical evolution over time, which is critical for accurately reconstructing earlier-stage distributions.

\section{Training results and discussion}
\label{results}

We utilize the comparison of the charge separation (CS) fraction between the U-Net predictions and the ground truth values, which refer to the CS fraction for different times or stages of HIC evolution generated by the AMPT model, as a key metric to evaluate the accuracy of the predicted reverse evolution for our training. The prediction accuracy of event-average CS fraction at each time step is defined as,
\begin{equation}\label{accuracyeq}
Accuracy = 1 - \frac{|f_{\text{prediction}} - f_{\text{ground}\;\text{truth}}|}{f_{\text{ground}\;\text{truth}}}.
\end{equation} 
Three different cases have been investigated on the basis of the choice of input data, which is listed in Table~\ref{tab:model_configs}. In Case 1, the model is trained by using parton transverse momentum space distributions at the freeze-out stage as the input, while in Case 2 and Case 3, the transverse momentum space distributions of final-state hadron are used as the training input. In first two cases, the initial CS fraction is fixed at $f = 10 \% $. We extend the analysis by considering varying CS strengths in Case 3, with $f$ ranging from $1 \%$ to $10\% $ in increments of $1 \%$, allowing us to improve the model's robustness by different initial strengths of CS fraction. For each value of fraction $f$, 500 training events are generated, assuming equal probability across all values.

\begin{table*}[!htbp]
\centering
\caption{Three configuration cases classified by initial CME conditions in AMPT and input types to the U-Net.}
\label{tab:model_configs} 
\renewcommand{\arraystretch}{1.5} 
\begin{tabular}{c@{\hspace{0.5cm}}c@{\hspace{0.5cm}}c} 
\hline
\hline
\textbf{Case} & \textbf{Initial CME Condition} & \textbf{Input to U-Net} \\
\hline
Case 1 & Fixed $f=10\%$ & $p_T$ distribution of partons at QGP freeze-out  \\
Case 2 & Fixed $f=10\%$ & $p_T$ distribution of hadrons at hadron freeze-out \\
Case 3 & Mixed $f=1\%, 2\%, 3\%,...,10\%$ & $p_T$ distribution of hadrons at hadron freeze-out \\
\hline
\hline
\end{tabular}

\end{table*}

\subsection{Case 1: Results for the Partonic Phase Inversion }
\label{case1}
\begin{figure}[hbtp]
    \begin{flushleft}
	\includegraphics[scale=0.35]{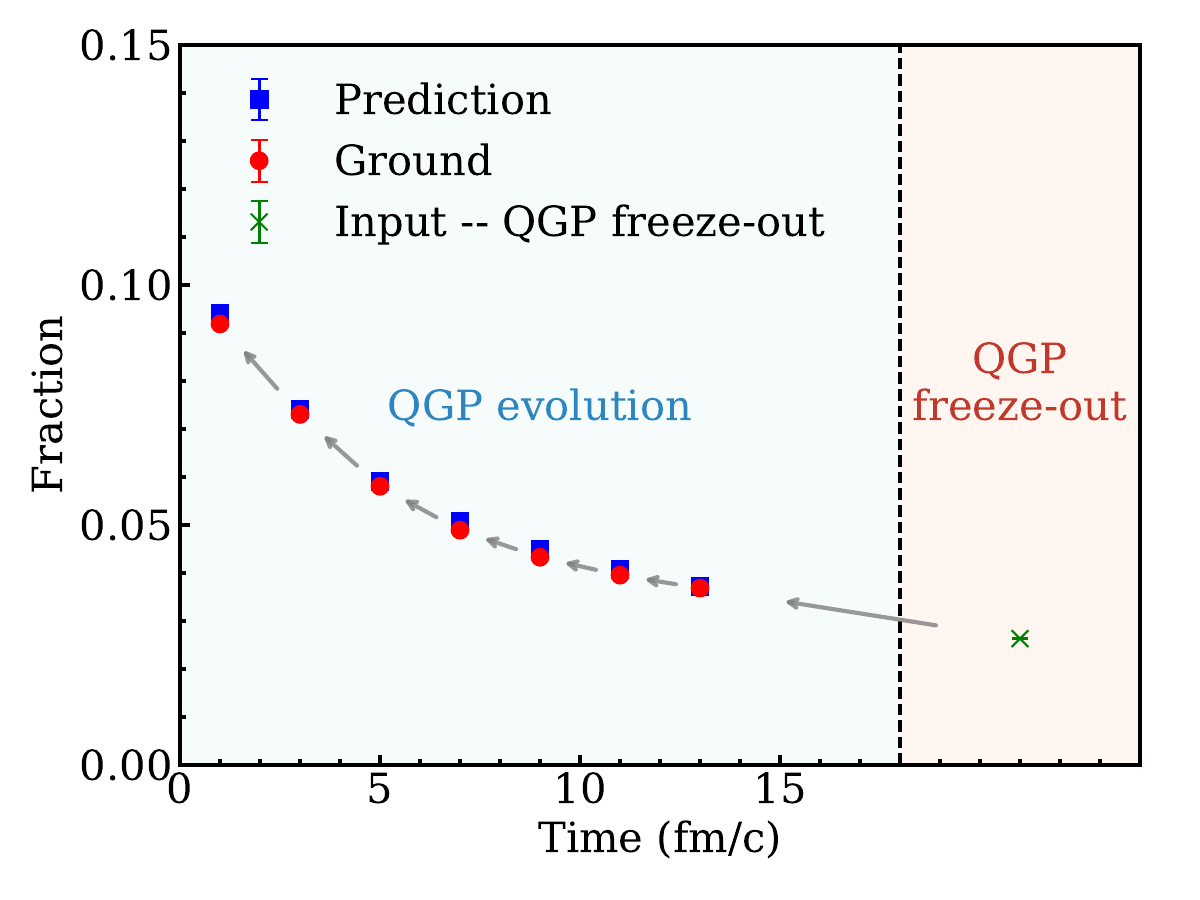}
	\caption{The inverse evolution of the charge separation (CS) fraction in the QGP phase. The star symbol (most right) indicates the input CS fraction at the freeze-out of QGP phase, while circular and square markers represent the ground truth and model predictions, respectively, at various early time points of QGP evolution. The gray arrows indicate the direction of the model's backward prediction, starting from the QGP freeze-out stage and tracing back to the initial QGP step by step in time.}
	\label{parton_event_average}
    \end{flushleft}
\end{figure}

\begin{figure}[hbtp]
    \begin{flushleft}
	\includegraphics[scale=0.35]{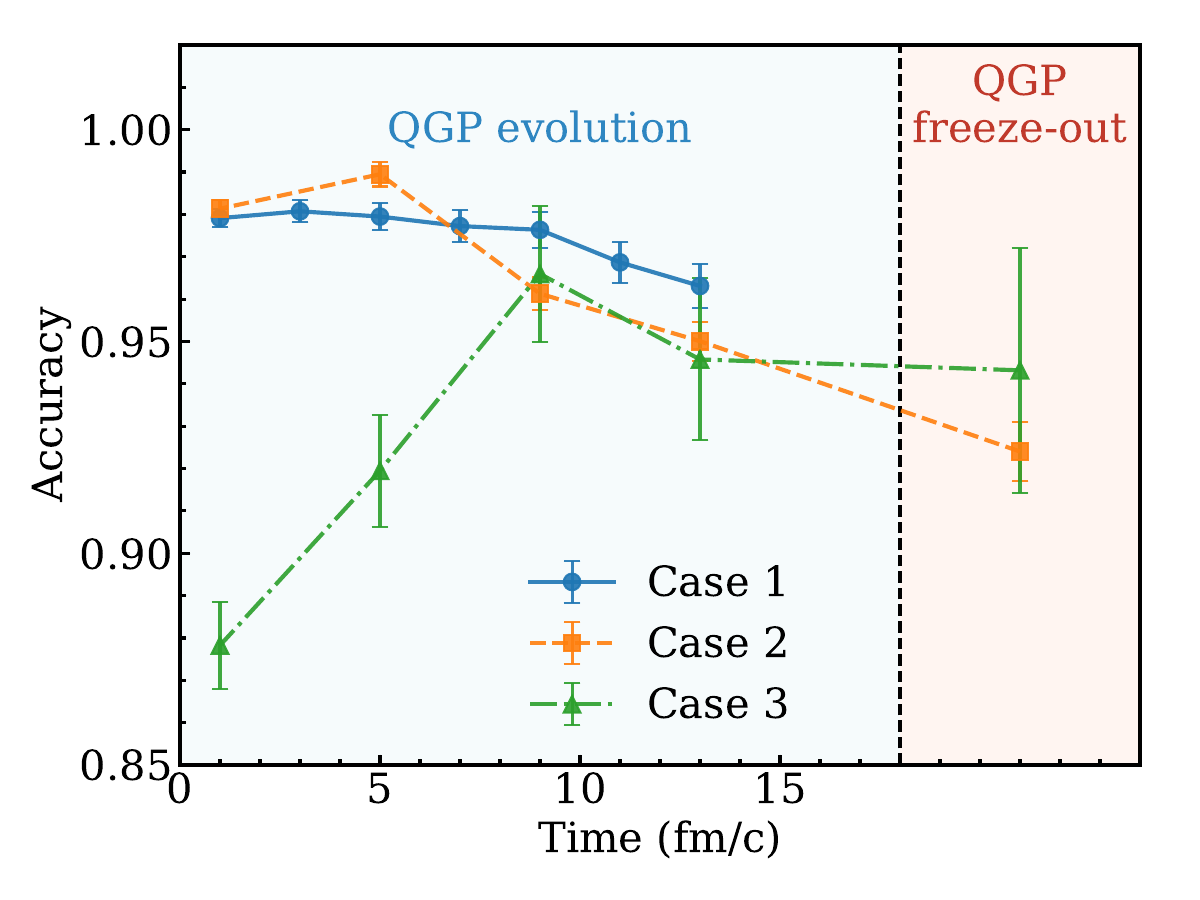}
	\caption{Prediction accuracy of event-average charge separation (CS) fraction at each time step for the three model configuration cases in Table~\ref{tab:model_configs}, where blue, orange, and green symbols represent Case 1, Case 2, and Case 3, respectively.}
	\label{accuarcy}
    \end{flushleft}
\end{figure}

Figure~\ref{parton_event_average}  illustrates the inverse evolution of charge separation (CS) fraction during the QGP phase (from right to left) on an event-averaged basis based on the test dataset. The star symbol on the most right corresponds to the input CS at the freeze-out of QGP phase, while the circular and square markers represent the ground truth and U-Net predictions, respectively. The CS fraction is computed from the transverse momentum distribution using Eq.~\eqref{charge_separation}. It is evident that the CS fraction decreases during the QGP evolution, dropping from approximately 10\% initially to about 3\% at the QGP freeze-out stage, primarily due to partonic interactions. Despite the significant degradation of the CS signal, the predicted event-averaged CS remains in good agreement with the ground truth, achieving an accuracy exceeding 96\% at each time step, as illustrated by the blue circles in Figure~\ref{accuarcy}.  As the CS evolves reversely, the prediction accuracy gradually improves, since information from later distributions contributes to better inference of the earlier stages.
 Notably, the model is able to infer the initial CS from a weak final signal, demonstrating its ability to reconstruct both the decreasing trend and the early-stage magnitude of CS throughout the QGP evolution, given the parton transverse momentum distribution at QGP freeze-out stage.

\begin{figure*}[hbtp]
	\includegraphics[scale=0.25]{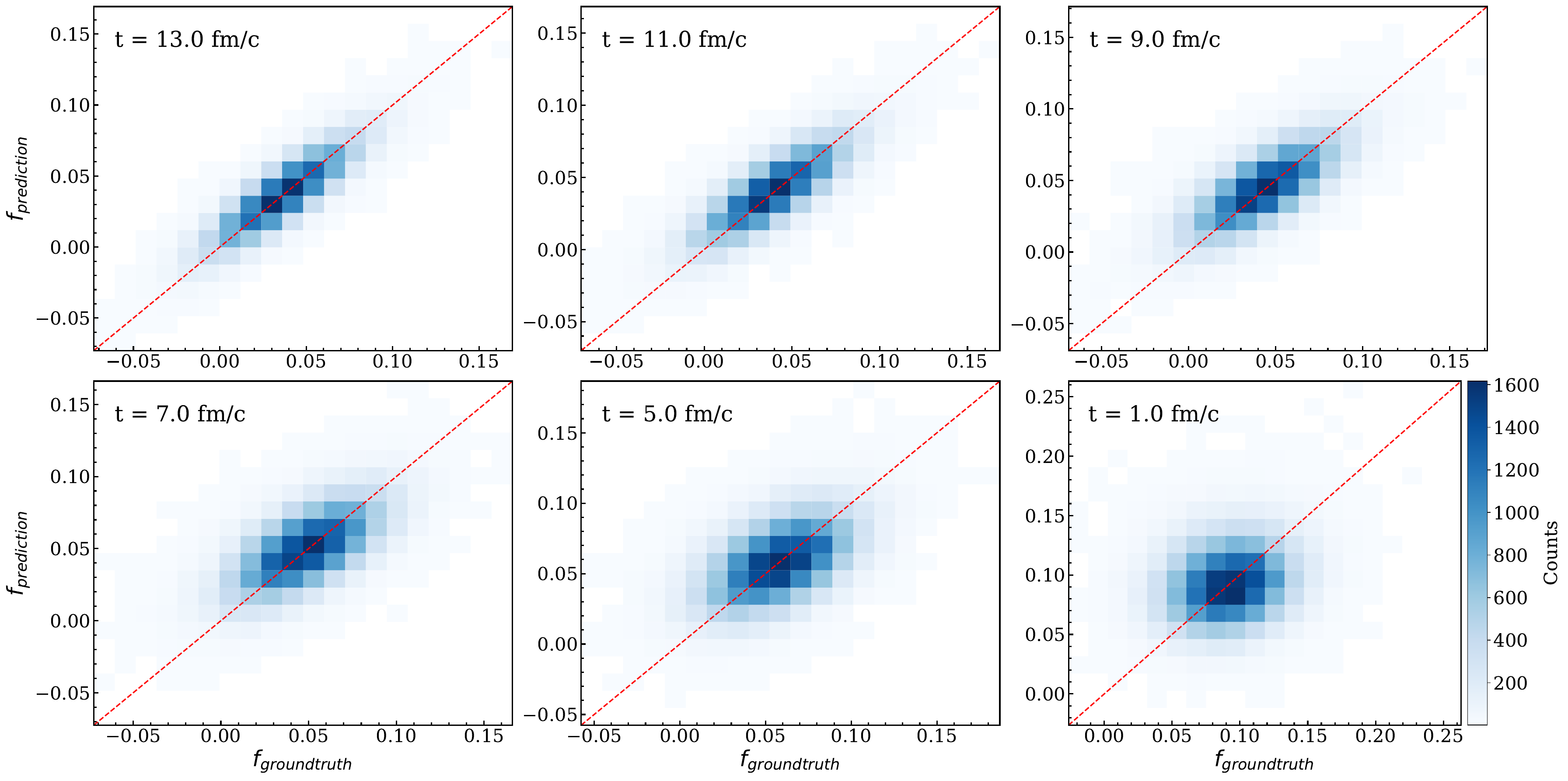}
	\caption{Two-dimensional histograms comparing model predictions with ground truth values for the charge separation(CS) fraction at selected time points: $t = 1.0, 5.0, 7.0, 9.0, 11.0, 13.0\,\mathrm{fm}/c$. The red dashed line ($y = x$) represents ideal agreement between predictions and ground truth values.}
	\label{parton_event_by_event}
\end{figure*}

Furthermore, Figure~\ref{parton_event_by_event}  presents a comparison between the event-by-event predictions of charge separation (CS) $f$ and the ground truth at different evolution times. The subfigures illustrate sequential time points, starting from the late QGP phase and tracing back to the early stage at $t=1\,\mathrm{fm}/c$. The strong linear correlation observed across the first several subfigures confirms the model's ability to capture the CS dynamics in the later stage of QGP phase on an event-by-event basis. However, we observe a gradual decline in prediction accuracy when evolving backward in time, which can be attributed to the large number of particles generated during the early-stage parton dynamics, as well as to error accumulation across recurrent prediction steps. One-dimensional histograms, comparing the distributions of the CS fractions predicted by the model and those from the ground truth at various QGP evolution time steps, are shown in the left column of Figure S1 in the Supplementary Material (SM). At each time point, the significant overlap between the predicted and true distributions highlights the effectiveness of the model in capturing the overall statistical behavior of the CS dynamics. This consistency across time steps further supports the ability of the model to reliably recover the temporal evolution of the CS fraction throughout the entire QGP phase, despite the complexity of the QGP dynamical evolution.

\subsection{Case 2: Results for the Hadronic Phase Inversion }
\label{case2}
\begin{figure}[hbtp]
    \begin{flushleft}
	\includegraphics[scale=0.34]{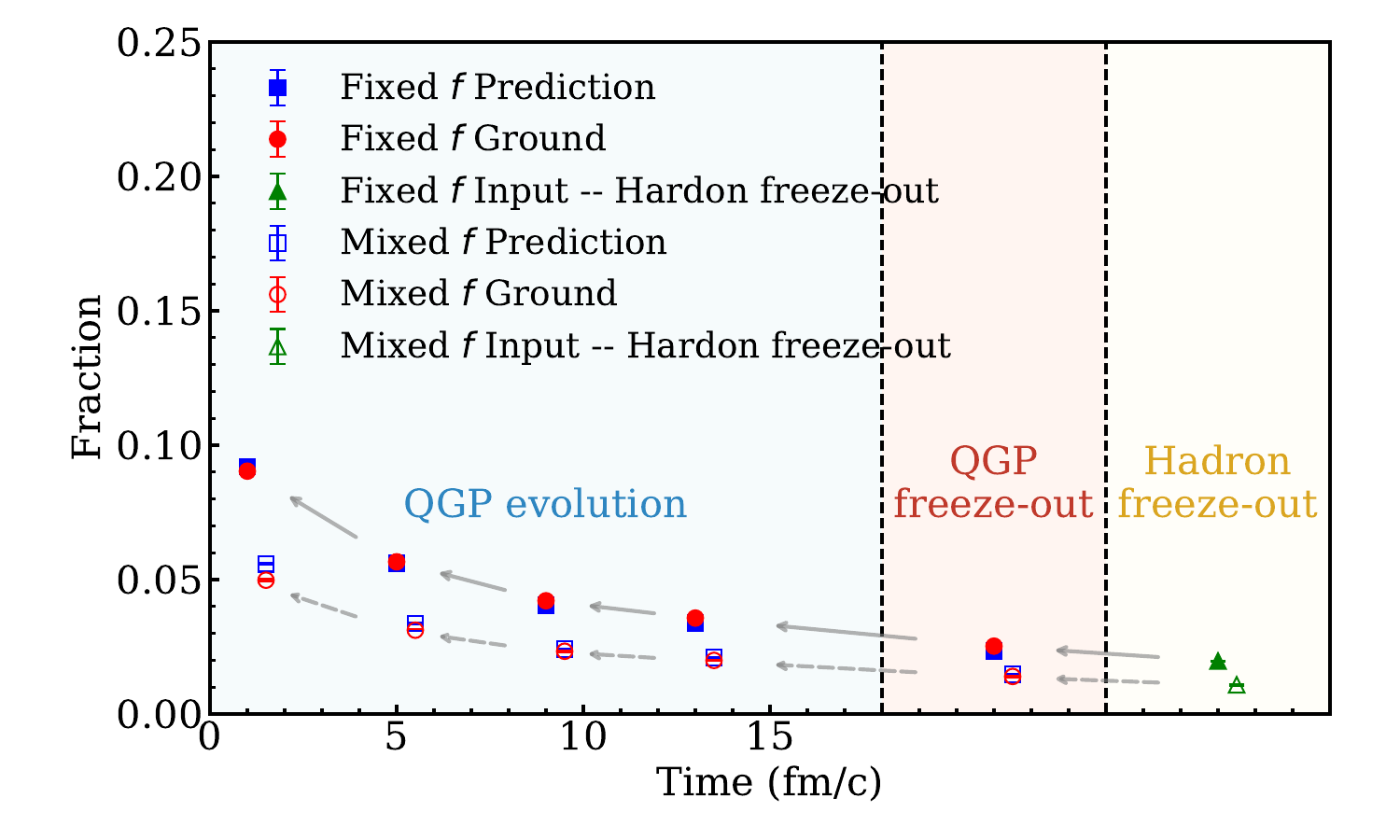}
	\caption{ The inverse evolution of the charge separation (CS) fraction across all stages of heavy-ion collisions. The triangular symbol (most right) indicates the input CS fraction from final-state hadron transverse momentum space, while circular and square markers represent the ground truth and model predictions, respectively, at the freeze-out of QGP phase, and various early time points of QGP evolution. Solid symbols correspond to the case 2 with fixed $f$ as described in Section ~\ref{case2}, whereas hollow symbols represent the case 3 of the mixed $f$ scenario discussed in Section ~\ref{case3}. The gray arrows indicate the backward time direction of the model  prediction. }
	\label{hadron_event_average}
    \end{flushleft}
\end{figure}

To achieve practical applicability of our model for experimental data analysis, we utilize the transverse momentum distribution of final-state hadrons as input. As shown by the solid symbols in Figure~\ref{hadron_event_average}, the ground truth and model predictions are represented by circular and square markers, respectively. The second rightmost point corresponds to the charge separation (CS) fraction at the QGP freeze-out. By moving leftward, we trace earlier stages of the QGP phase and recover the temporal evolution of CS from hadronic final state. Our model successfully time-reversely recovers the CS along the HICs evolution: beginning with final-state hadronic transverse momentum distribution, it first predicts the distribution at QGP freeze-out stage, then successfully simulates the reverse evolution to earlier partonic stages. Remarkably, despite the complexity of the hadronization process, the prediction accuracy of CS reconstruction gradually increases during the reverse evolution, reaching over 97\% at the initial QGP stage, as indicated by the orange markers in Figure~\ref{accuarcy}, which demonstrates its strong capability to reconstruct the CS dynamics throughout the full timeline of HICs evolution.

The comparative one-dimensional histograms of the model predictions versus ground truth charge separation (CS) values at each evolution stage are presented in the middle column of Figure S1 in the Supplementary Material (SM). The results show the consistency between the model predictions and ground truth throughout all evolution stage, with distribution overlap exceeding 95$\%$ at each time point. This agreement indicates that, given the final-state hadron distribution as input, our model can predict the mean value of CS and its distribution in the test dataset at an arbitrary time point of QGP evolution.

\subsection{Case 3: Results for mixed charge separation (CS) strength }
\label{case3}

To evaluate the robustness of our model in capturing the charge separation (CS) dynamics under varying initial conditions, we generate a series of initial states with different strengths of CS, ranging from $1 \%$ to $10\% $ in increments of $1 \%$. In this case, we present results from a model trained on data with varying initial conditions, using final-state hadrons as input, a setup that is more practical for experimental applications.

The hollow symbols in Figure~\ref{hadron_event_average} represent the reverse evolution of the event-averaged charge separation (CS) fraction in this case. Throughout the HICs evolution, the CS fraction decreases from approximately 5\% in the initial QGP stage to about 1\% at hadron freeze-out. Inputting transverse momentum distribution of the final-state hadrons with weak CME-induced signal, the model is still able to reconstruct the CME signal throughout the entire evolution, achieving a minimum accuracy of 87\% at the earliest time step, as indicated by the green markers in Figure~\ref{accuarcy}. Furthermore, the right column of Figure S1 in the Supplementary Material (SM)provides a detailed comparison between the predicted CS distributions and the ground truth at each time step. These results demonstrate that even with varying initial CS strengths, amplifying fluctuations, and using only final-state hadrons as input, our model can still robustly reconstruct the event-average CS evolution and reproduce its distribution at any intermediate time of the QGP evolution. This robustness highlights the ability of our model to extract the underlying CS dynamics from final-state observables, making it a powerful tool for studying the properties of the CME in HICs.

\section{Summary and outlook}
\label{summary} 
In summary, we adopt a deep learning approach to reconstruct the CME-like signal implemented in AMPT by learning its reverse dynamic evolution. Employing a time-embedded U-Net architecture, our model successfully captures both the trend and magnitude of charge separation throughout the HICs evolution. When tested on AMPT simulations with fixed initial charge separation fraction, the method accurately predicts both the event-averaged charge separation along the reverse evolution of HICs as well as the charge separation distribution at each time step. Notably, the approach remains robust with a wide range of initial charge separation strengths, demonstrating its effectiveness in reconstructing the early-state charge separation signals from experimentally accessible final-state transverse momentum distribution. This Letter thus establishes an advanced methodological framework based on deep learning, capable of time-reversely unfolding signal dynamics and opening a novel perspective for investigating the fundamental properties of the QGP and other related physics in HICs.

To further enhance the feasibility and applicability of the method, particularly under realistic experimental conditions, we propose two key directions for future improvement. First, in simulation models, we will extend the current framework by incorporating insights from chiral kinetic theory, a well-established formalism for describing the non-equilibrium dynamics of chiral systems, which has been widely used to study chiral magnetic and vortical effects in HIC \cite{CKT_1,CKT_2,CKT_3,CKT_Sun_1,CKT_Sun_2,CKT_Xu_1,CKT_Xu_2}. Adopting this framework provides a stronger physical foundation and will allow us to capture the more realistic time-reversal evolution of chiral anomalous processes. A particular focus will be placed on the Chiral Anomaly Transport (CAT) model \cite{CAT_1,CAT_2}, which utilizes chiral kinetic theory to dynamically generate charge separation along the magnetic field in the presence of chirality imbalance. By integrating training data from such realistic simulations, the proposed deep-learning framework is expected to directly link the unfolded early-time charge separation to its dynamical origin. This capability will provide a novel and model-informed approach to quantify key characteristics of the chiral magnetic effect in HICs. Second, we aim to improve the deep learning architecture by integrating PointNet~\cite{pointnet1}, a neural network designed for directly processing point cloud data. PointNet enables high-dimensional data handling and advanced feature extraction, and has already shown promise in high-energy nuclear physics applications~\cite{pointnet2,pointnet3,pointnet4}. With these advancements, the model is expected to more accurately capture the complex and fluctuating dynamics of CME signals.

\begin{acknowledgments}
We thank Prof. Mei Huang and Dr. Zilin Yuan for providing the CAT model for our future derivative work. This work is partially supported by the National Natural Science Foundation of China under Grants No.12147101 and No. 12325507, the National Key Research and Development Program of China under Grant No. 2022YFA1604900, the Guangdong Major Project of Basic and Applied Basic Research under Grant No. 2020B0301030008 (S.G. and G.M.), the CUHK-Shenzhen university development fund under grant No.\ UDF01003041 and UDF03003041, and Shenzhen Peacock fund under No.\ 2023TC0179 (K.Z.), the RIKEN TRIP initiative (RIKEN Quantum), JSPS KAKENHI Grant No. 25H01560, and JST-BOOST Grant No.JPMJBY24H9 (L.W.). We also thank the DEEP-IN working group at RIKEN-iTHEMS for support in the preparation of this Letter.

\end{acknowledgments}


\begin{thebibliography}{99}


\bibitem {QCD1} Lee~T~D and Wick~G~C \href{https://doi.org/10.1103/PhysRevD.9.2291}{1974 {\it Phys. Rev. D} {\bf 9} 2291}

\bibitem {QCD2} Kharzeev~D, Pisarski~R~D, and Tytgat~M~H~G \href{https://doi.org/10.1103/PhysRevLett.81.512}{1998 {\it Phys. Rev. Lett.} {\bf 81} 512}

\bibitem {QCD3} Kharzeev~D and Pisarski~R~D \href{https://doi.org/10.1103/PhysRevD.61.111901}{2000 {\it Phys. Rev. D} {\bf 61} 111901}

\bibitem {CME1} Kharzeev~D~E, Liao~J, Voloshin~S~A, and Wang~G \href{https://doi.org/10.1016/j.ppnp.2016.01.001}{2016 {\it Prog. Part. Nucl. Phys.} {\bf 88} 1}

\bibitem {CME2} Fukushima~K, Kharzeev~D~E, and Warringa~H~J \href{https://doi.org/10.1103/PhysRevD.78.074033}{2008 {\it Phys. Rev. D} {\bf 78} 074033}

\bibitem {CME3} Kharzeev~D \href{https://doi.org/10.1016/j.physletb.2005.11.075}{2006 {\it Phys. Lett. B} {\bf 633} 260}

\bibitem {CME4} Kharzeev~D and Zhitnitsky~A \href{https://doi.org/10.1016/j.nuclphysa.2007.10.001}{2007 {\it Nucl. Phys. A} {\bf 797} 67}

\bibitem {CME5} Hattori~K and Huang~X~G \href{https://doi.org/10.1007/s41365-016-0178-3}{2017 {\it Nucl. Sci. Tech.} {\bf 28} 26}


\bibitem {magentic} Skokov~V~V, Illarionov~A~Y, and Toneev~V~D \href{https://doi.org/10.1142/S0217751X09047570}{2009 {\it Int. J. Mod. Phys. A} {\bf 24} 5925}

\bibitem {Bzdak:2011yy} Bzdak~A and Skokov~V \href{https://doi.org/10.1016/j.physletb.2012.02.065}{2012 {\it Phys. Lett. B} {\bf 710} 171}

\bibitem {early_stage} Deng~W~T and Huang~X~G \href{https://doi.org/10.1103/PhysRevC.85.044907}{2012 {\it Phys. Rev. C} {\bf 85} 044907}

\bibitem {Chen:2021nxs} Chen~Y, Sheng~X~L, and Ma~G~L \href{https://doi.org/10.1016/j.nuclphysa.2021.122199}{2021 {\it Nucl. Phys. A} {\bf 1011} 122199}


\bibitem {Wu:2021xgu} Wu~S, Shen~C, and Song~H \href{https://doi.org/10.1088/0256-307X/38/8/081201}{2021 {\it Chin. Phys. Lett.} {\bf 38} 081201}



\bibitem {Zhao:2017rpf} Zhao~X~L, Ma~Y~G, and Ma~G~L \href{https://doi.org/10.1103/PhysRevC.97.024910}{2018 {\it Phys. Rev. C} {\bf 97} 024910}


\bibitem {Zhao:2022dac} Zhao~J, Chen~J~H, Huang~X~G, and Ma~Y~G \href{https://doi.org/10.1007/s41365-024-01374-9}{2024 {\it Nucl. Sci. Tech.} {\bf 35} 20}


\bibitem {Shou:2024uga} Shou~Q~Y, Ma~Y~G, Zhang~S, Zhu~J~H, Mao~Y~X, Pei~H, Yin~Z~B, Zhang~X~M, Zhou~D~C, Peng~X~Y, Bai~X~Z, Tang~Z~B, Zhang~Y~F, and Li~X~M \href{https://doi.org/10.1007/s41365-024-01583-2}{2024 {\it Nucl. Sci. Tech.} {\bf 35} 219}

\bibitem {Gao:2020vbh} Gao~J~H, Ma~G~L, Pu~S, and Wang~Q \href{https://doi.org/10.1007/s41365-020-00801-x}{2020 {\it Nucl. Sci. Tech.} {\bf 31} 90}

\bibitem {Xu:2014tda} Xu~J, Liao~J, and Gyulassy~M \href{https://doi.org/10.1088/0256-307X/32/9/092501}{2015 {\it Chin. Phys. Lett.} {\bf 32} 092501}

\bibitem {Tang:2011xq} Tang~Z~B, Yi~L, Ruan~L~J, Shao~M, Chen~H~F, Li~C, Mohanty~B, and Xu~Z~B \href{https://doi.org/10.1088/0256-307X/30/3/031201}{2013 {\it Chin. Phys. Lett.} {\bf 30} 031201}

\bibitem{Shen:2025unr} Shen~D~Y, Chen~J~H, Huang~X~G, Ma~Y~G, Tang~A~H, and Wang~G \href{https://doi.org/10.34133/research.0726}{2025 {\it Research} {\bf 8} 0726}

\bibitem {RHIC1} Abelev~B~I, Adam~J, Adare~A~M {\it et al.} (STAR Collaboration) \href{https://doi.org/10.1103/PhysRevLett.103.251601}{2009 {\it Phys. Rev. Lett.} {\bf 103} 251601}

\bibitem {RHIC2} Abelev~B~I, Aggarwal~M~M, Ahammed~Z {\it et al.} (STAR Collaboration) \href{https://doi.org/10.1103/PhysRevC.81.054908}{2010 {\it Phys. Rev. C} {\bf 81} 054908}

\bibitem {RHIC3} Adamczyk~L, Adkins~J~K, Agakishiev~G {\it et al.} (STAR Collaboration) \href{https://doi.org/10.1103/PhysRevLett.113.052302}{2014 {\it Phys. Rev. Lett.} {\bf 113} 052302}

\bibitem{Chen:2024aom} Chen~J~H {\it et al.} \href{https://doi.org/10.1007/s41365-024-01591-2}{2024 {\it Nucl. Sci. Tech.} {\bf 35} 214}

\bibitem {LHC1} Abelev~B, Adam~J, Adamova~D {\it et al.} (ALICE Collaboration) \href{https://doi.org/10.1103/PhysRevLett.110.012301}{2013 {\it Phys. Rev. Lett.} {\bf 110} 012301}

\bibitem {LHC2} Acharya~S, Adam~J, Adamova~D {\it et al.} (ALICE Collaboration) \href{https://doi.org/10.1016/j.physletb.2017.12.021}{2018 {\it Phys. Lett. B} {\bf 777} 151}

\bibitem {background1} Adamczyk~L, Adkins~J~K, Agakishiev~G {\it et al.} (STAR Collaboration) \href{https://doi.org/10.1103/PhysRevC.89.044908}{2014 {\it Phys. Rev. C} {\bf 89} 044908}

\bibitem {background2} Wang~F and Zhao~J \href{https://doi.org/10.1103/PhysRevC.95.051901}{2017 {\it Phys. Rev. C} {\bf 95} 051901}

\bibitem {background3} Zhao~J (STAR Collaboration) \href{https://doi.org/10.1016/j.nuclphysa.2018.08.035}{2019 {\it Nucl. Phys. A} {\bf 982} 535}

\bibitem {background4} Feng~Y, Voloshin~S~A, and Wang~F \href{https://arxiv.org/abs/2502.09742}{2025 {\it arXiv:2502.09742}}

\bibitem {CME_analysis1} Sirunyan~A~M, Tumasyan1~A, Adam~W {\it et al.} (CMS Collaboration) \href{https://doi.org/10.1103/PhysRevC.97.044912}{2018 {\it Phys. Rev. C} {\bf 97} 044912}

\bibitem {CME_analysis2} Xu~H~J, Zhao~J, Wang~X, Li~H, Lin~Z~W, Shen~C, and Wang~F \href{https://doi.org/10.1088/1674-1137/42/8/084103}{2018 {\it Chin. Phys. C} {\bf 42} 084103}

\bibitem {CME_analysis3} Tang~A~H \href{https://doi.org/10.1088/1674-1137/44/5/054101}{2020 {\it Chin. Phys. C} {\bf 44} 054101}

\bibitem {isobar1} Koch~V, Schlichting~S, Skokov~V, Sorensen~P, Thomas~J, Voloshin~S, Wang~G, and Yee~H~U \href{https://doi.org/10.1088/1674-1137/41/7/072001}{2017 {\it Chin. Phys. C} {\bf 41} 072001}

\bibitem {isobar2} Abdallah~M~S, Aboona~B~E, Adam~J {\it et al.} (STAR Collaboration) \href{https://doi.org/10.1103/PhysRevC.105.014901}{2022 {\it Phys. Rev. C} {\bf 105} 014901}


\bibitem {background_nuclear_structure1} Li~H, Xu~H~J, Zhao~J, Lin~Z~W, Zhang~H, Wang~X, Shen~C, and Wang~F \href{https://doi.org/10.1103/PhysRevC.98.054907}{2018 {\it Phys. Rev. C} {\bf 98} 054907}

\bibitem {background_nuclear_structure2} Zhao~X~L, Ma~G~L, and Ma~Y~G \href{https://doi.org/10.1103/PhysRevC.99.034903}{2019 {\it Phys. Rev. C} {\bf 99} 034903}

\bibitem {background_nuclear_structure3} Xu~H~J, Li~H, Wang~X, Shen~C, and Wang~F \href{https://doi.org/10.1016/j.physletb.2021.136453}{2021 {\it Phys. Lett. B} {\bf 819} 136453}

\bibitem {event_shape_engneering} Schukraft~J, Timmins~A, and Voloshin~S~A \href{https://doi.org/10.1016/j.physletb.2013.01.045}{2013 {\it Phys. Lett. B} {\bf 719} 394}

\bibitem {event_shape_selection1} STAR Collaboration \href{https://arxiv.org/abs/2506.00275}{2025 {\it arXiv:2506.00275}}

\bibitem {event_shape_selection2} STAR Collaboration \href{https://arxiv.org/abs/2506.00278}{2025 {\it arXiv:2506.00278}}

\bibitem {spectator_participant_plane1} Abdallah~M~S, Adam~J, Adamczyk~L {\it et al.} (STAR Collaboration) \href{https://doi.org/10.1103/PhysRevLett.128.092301}{2022 {\it Phys. Rev. Lett.} {\bf 128} 092301}

\bibitem {spectator_participant_plane2} Feng~Y, Zhao~J, Li~H, Xu~H~J, and Wang~F \href{https://doi.org/10.1103/PhysRevC.105.024913}{2022 {\it Phys. Rev. C} {\bf 105} 024913}

\bibitem {Chen:2025jmb} Chen~B~X, Zhao~X~L, and Ma~G~L \href{https://doi.org/10.1007/s41365-025-01761-w}{2025 {\it Nucl. Sci. Tech.} {\bf 36} 170}

\bibitem {Chen:2023jhx} Chen~B~X, Zhao~X~L, and Ma~G~L \href{https://doi.org/10.1103/PhysRevC.109.024909}{2024 {\it Phys. Rev. C} {\bf 109} 024909}

\bibitem {DL1} Schmidhuber~J \href{https://doi.org/10.1016/j.neunet.2014.09.003}{2015 {\it Neural Networks} {\bf 61} 85}

\bibitem {DL2} LeCun~Y, Bengio~Y, and Hinton~G \href{https://doi.org/10.1038/nature14539}{2015 {\it Nature} {\bf 521} 436}

\bibitem {Zhou:2023pti} Zhou~K, Wang~L, Pang~L~G, and Shi~S \href{https://doi.org/10.1016/j.ppnp.2023.104084}{2024 {\it Prog. Part. Nucl. Phys.} {\bf 135} 104084}

\bibitem {DL_nuclear1} Pang~L~G, Zhou~K, Su~N, Petersen~H, St{\"o}cker~H, and Wang~X~N \href{https://doi.org/10.1038/s41467-017-02726-3}{2018 {\it Nature Commun.} {\bf 9} 210}

\bibitem {DL_nuclear2} Zhou~K, Endr{\H{o}}di~G, Pang~L~G, and St{\"o}cker~H \href{https://doi.org/10.1103/PhysRevD.100.011501}{2019 {\it Phys. Rev. D} {\bf 100} 011501}

\bibitem {DL_nuclear3} Liu~Z, Zhao~W, and Song~H \href{https://doi.org/10.1140/epjc/s10052-019-7379-y}{2019 {\it Eur. Phys. J. C} {\bf 79} 870}

\bibitem {DL_nuclear4} Huang~H, Xiao~B, Liu~Z, Wu~Z, Mu~Y, and Song~H \href{https://doi.org/10.1103/PhysRevResearch.3.023256}{2021 {\it Phys. Rev. Res.} {\bf 3} 023256}

\bibitem {DL_nuclear5} Du~Y~L, Zhou~K, Steinheimer~J, Pang~L~G, Motornenko~A, Zong~H~S, Wang~X~N, and St{\"o}cker~H \href{https://doi.org/10.1140/epjc/s10052-020-8030-7}{2020 {\it Eur. Phys. J. C} {\bf 80} 516}

\bibitem {DL_nuclear7} Jiang~L, Wang~L, and Zhou~K \href{https://doi.org/10.1103/PhysRevD.103.116023}{2021 {\it Phys. Rev. D} {\bf 103} 116023}

\bibitem {DL_nuclear8} Shi~S, Zhou~K, Zhao~J, Mukherjee~S, and Zhuang~P \href{https://doi.org/10.1103/PhysRevD.105.014017}{2022 {\it Phys. Rev. D} {\bf 105} 014017}

\bibitem {DL_nuclear9} Wang~L, Shi~S, and Zhou~K \href{https://doi.org/10.1103/PhysRevD.106.L051502}{2022 {\it Phys. Rev. D} {\bf 106} L051502}

\bibitem {DL_nuclear10} Aarts~G, Fukushima~K, Hatsuda~T, Ipp~A, Shi~S, Wang~L, and Zhou~K \href{https://doi.org/10.1038/s42254-024-00798-x}{2025 {\it Nature Rev. Phys.} {\bf 7} 154}

\bibitem {pointnet2} Omana Kuttan~M, Steinheimer~J, Zhou~K, Redelbach~A, and Stoecker~H \href{https://doi.org/10.1016/j.physletb.2020.135872}{2020 {\it Phys. Lett. B} {\bf 811} 135872}

\bibitem {DL_particle1} Baldi~P, Sadowski~P, and Whiteson~D \href{https://doi.org/10.1038/ncomms5308}{2014 {\it Nature Commun.} {\bf 5} 4308}

\bibitem {DL_particle2} Baldi~P, Sadowski~P, and Whiteson~D \href{https://doi.org/10.1103/PhysRevLett.114.111801}{2015 {\it Phys. Rev. Lett.} {\bf 114} 111801}

\bibitem {DL_particle3} Barnard~J, Dawe~E~N, Dolan~M~J, and Rajcic~N \href{https://doi.org/10.1103/PhysRevD.95.014018}{2017 {\it Phys. Rev. D} {\bf 95} 014018}

\bibitem {DL_particle4} Broecker~P, Carrasquilla~J, Melko~R~G, and Trebst~S \href{https://doi.org/10.1038/s41598-017-09098-0}{2017 {\it Sci. Rep.} {\bf 7} 8823}

\bibitem {DL_particle5} Radovic~A, Williams~M, Rousseau~D, Kagan~M, Bonacorsi~D, Himmel~A, Aurisano~A, Terao~K, and Wongjirad~T \href{https://doi.org/10.1038/s41586-018-0361-2}{2018 {\it Nature} {\bf 560} 41}

\bibitem {DL_condensed1} Carrasquilla~J and Melko~R~G \href{https://doi.org/10.1038/nphys4035}{2017 {\it Nature Phys.} {\bf 13} 431}

\bibitem {DL_condensed2} van Nieuwenburg~E~P~L, Liu~Y~H, and Huber~S~D \href{https://doi.org/10.1038/nphys4037}{2017 {\it Nature Phys.} {\bf 13} 435}

\bibitem {DL_condensed3} Han~Z~Y, Wang~J, Fan~H, Wang~L, and Zhang~P \href{https://doi.org/10.1103/PhysRevX.8.031012}{2018 {\it Phys. Rev. X} {\bf 8} 031012}

\bibitem {DL_condensed4} Carleo~G, Cirac~I, Cranmer~K, Daudet~L, Schuld~M, Tishby~N, Vogt-Maranto~L, and Zdeborov{\'a}~L \href{https://doi.org/10.1103/RevModPhys.91.045002}{2019 {\it Rev. Mod. Phys.} {\bf 91} 045002}

\bibitem {CME_meter} Zhao~Y~S, Wang~L, Zhou~K, and Huang~X~G \href{https://doi.org/10.1103/PhysRevC.106.L051901}{2022 {\it Phys. Rev. C} {\bf 106} L051901}

\bibitem {AMPT_CS} Ma~G~L and Zhang~B \href{https://doi.org/10.1016/j.physletb.2011.04.057}{2011 {\it Phys. Lett. B} {\bf 700} 39}

\bibitem {CKT_Sun_1} Sun~Y, Ko~C~M, and Li~F \href{https://doi.org/10.1103/PhysRevC.94.045204}{2016 {\it Phys. Rev. C} {\bf 94} 045204}

\bibitem {AVFD_1} Shi~S, Jiang~Y, Lilleskov~E, and Liao~J \href{https://doi.org/10.1016/j.aop.2018.04.026}{2018 {\it Annals Phys.} {\bf 394} 50}

\bibitem {AMPT} Lin~Z~W, Ko~C~M, Li~B~A, Zhang~B, and Pal~S \href{https://doi.org/10.1103/PhysRevC.72.064901}{2005 {\it Phys. Rev. C} {\bf 72} 064901}

\bibitem {HIJING1} Wang~X~N and Gyulassy~M \href{https://doi.org/10.1103/PhysRevD.44.3501}{1991 {\it Phys. Rev. D} {\bf 44} 3501}

\bibitem {HIJING2} Gyulassy~M and Wang~X~N \href{https://doi.org/10.1016/0010-4655(94)90057-4}{1994 {\it Comput. Phys. Commun.} {\bf 83} 307}

\bibitem {HIJING3} Lin~Z~W and Ko~C~M \href{https://doi.org/10.1103/PhysRevC.65.034904}{2002 {\it Phys. Rev. C} {\bf 65} 034904}

\bibitem {ZPC} Zhang~B \href{https://doi.org/10.1016/S0010-4655(98)00010-1}{1998 {\it Comput. Phys. Commun.} {\bf 109} 193}

\bibitem {3mb0} Lin~Z~W \href{https://doi.org/10.1103/PhysRevC.90.014904}{2014 {\it Phys. Rev. C} {\bf 90} 014904}

\bibitem {3mb1} Orjuela Koop~J~D, Adare~A, McGlinchey~D, and Nagle~J~L \href{https://doi.org/10.1103/PhysRevC.92.054903}{2015 {\it Phys. Rev. C} {\bf 92} 054903}

\bibitem {3mb2} Ma~G~L and Bzdak~A \href{https://doi.org/10.1016/j.nuclphysa.2016.01.057}{2016 {\it Nucl. Phys. A} {\bf 956} 745}

\bibitem {3mb3} Ma~G~L and Lin~Z~W \href{https://doi.org/10.1103/PhysRevC.93.054911}{2016 {\it Phys. Rev. C} {\bf 93} 054911}

\bibitem {3mb4} He~Y and Lin~Z~W \href{https://doi.org/10.1103/PhysRevC.96.014910}{2017 {\it Phys. Rev. C} {\bf 96} 014910}

\bibitem {3mb5} Lin~Z~W and Zheng~L \href{https://doi.org/10.1007/s41365-021-00944-5}{2021 {\it Nucl. Sci. Tech.} {\bf 32} 113}

\bibitem {ART} Li~B~A and Ko~C~M \href{https://doi.org/10.1103/PhysRevC.52.2037}{1995 {\it Phys. Rev. C} {\bf 52} 2037}



\bibitem {Zhao:2022grq} Zhao~X~L and Ma~G~L \href{https://doi.org/10.1103/PhysRevC.106.034909}{2022 {\it Phys. Rev. C} {\bf 106} 034909}


\bibitem {Unet_arti} Ronneberger~O, Fischer~P, and Brox~T \href{https://arxiv.org/abs/1505.04597}{2015 {\it arXiv:1505.04597}}

\bibitem {pt_cut} Abelev~B~I, Adams~J, Aggarwal~M~M {\it et al.} (STAR Collaboration) \href{https://doi.org/10.1103/PhysRevC.75.064901}{2007 {\it Phys. Rev. C} {\bf 75} 064901}

\bibitem {CKT_1} Son~D~T and Yamamoto~N \href{https://doi.org/10.1103/PhysRevLett.109.181602}{2012 {\it Phys. Rev. Lett.} {\bf 109} 181602}

\bibitem {CKT_2} Stephanov~M~A and Yin~Y \href{https://doi.org/10.1103/PhysRevLett.109.162001}{2012 {\it Phys. Rev. Lett.} {\bf 109} 162001}

\bibitem {CKT_3} Gao~J~H, Liang~Z~T, Pu~S, Wang~Q, and Wang~X~N \href{https://doi.org/10.1103/PhysRevLett.109.232301}{2012 {\it Phys. Rev. Lett.} {\bf 109} 232301}


\bibitem {CKT_Sun_2} Sun~Y and Ko~C~M \href{https://doi.org/10.1103/PhysRevC.98.014911}{2018 {\it Phys. Rev. C} {\bf 98} 014911}

\bibitem {CKT_Xu_1} Zhou~W~H and Xu~J \href{https://doi.org/10.1016/j.physletb.2019.134932}{2019 {\it Phys. Lett. B} {\bf 798} 134932}

\bibitem {CKT_Xu_2} Zhou~W~H and Xu~J \href{https://doi.org/10.1103/PhysRevC.98.044904}{2018 {\it Phys. Rev. C} {\bf 98} 044904}

\bibitem {CAT_1} Yuan~Z, Huang~A, Zhou~W~H, Ma~G~L, and Huang~M \href{https://doi.org/10.1103/PhysRevC.109.L031903}{2024 {\it Phys. Rev. C} {\bf 109} L031903}

\bibitem {CAT_2} Yuan~Z, Huang~A, Xie~G, Zhou~W~H, Ma~G~L, and Huang~M \href{https://doi.org/10.1103/PhysRevC.111.044913}{2025 {\it Phys. Rev. C} {\bf 111} 044913}

\bibitem {pointnet1} Qi~C~R, Su~H, Mo~K, and Guibas~L~J \href{https://arxiv.org/abs/1612.00593}{2016 {\it arXiv:1612.00593}}


\bibitem {pointnet3} Omana Kuttan~M, Steinheimer~J, Zhou~K, Redelbach~A, and Stoecker~H \href{https://doi.org/10.3390/particles4010006}{2021 {\it Particles} {\bf 4} 47}

\bibitem {pointnet4} Guo~S, Wang~H~S, Zhou~K, and Ma~G~L \href{https://doi.org/10.1103/PhysRevC.110.024910}{2024 {\it Phys. Rev. C} {\bf 110} 024910}





\hypersetup{urlcolor=red}
\end{thebibliography}

\clearpage

\begin{titlepage}
\centering
{\large\bfseries Supplemental Materials for \\[0.5cm]
``Neural Unfolding of the Chiral Magnetic Effect in Heavy-Ion Collisions''}
\maketitle
\end{titlepage}

\onecolumngrid

\subsection*{One-Dimensional Histograms of Charge Separation}

One-dimensional histograms comparing the charge separation (CS) fraction distributions between model predictions and ground truth values in three cases in section about training results and discussion are shown in Figure~\ref{hist}.

\setcounter{figure}{0}
\renewcommand{\thefigure}{S\arabic{figure}}
\begin{figure*}[hbtp]
	\includegraphics[scale=0.20]{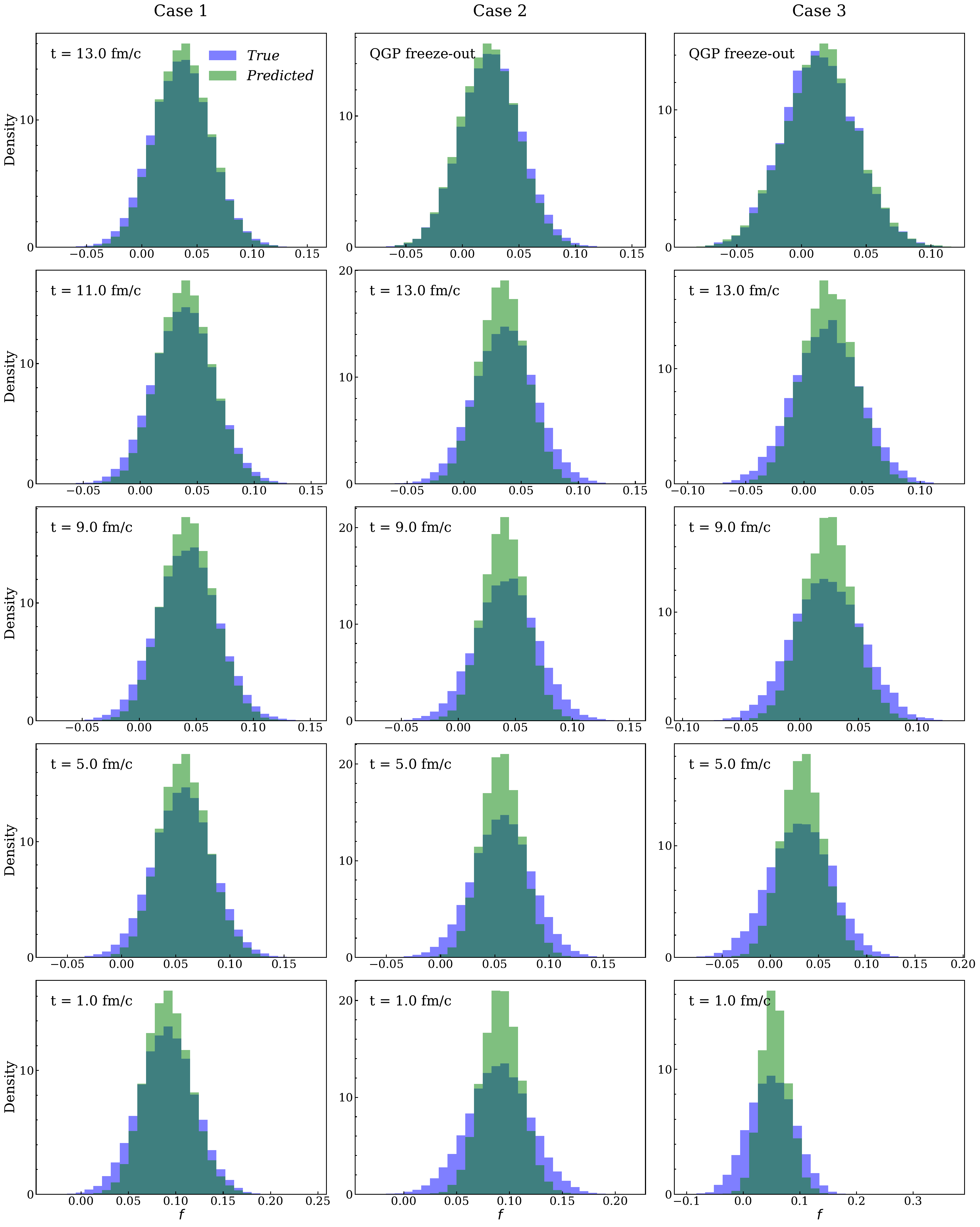}
	\caption{One-dimensional histograms of charge separation (CS) fractions comparing model predictions (green) and ground truth (blue) across three cases. Columns represent cases; rows represent different time steps.}
	\label{hist}
\end{figure*}
\end{CJK*}
\end{document}